\documentclass{llncs}
\pdfoutput=1
\usepackage{makeidx}
\usepackage[utf8]{inputenc}
\usepackage{graphicx}
\usepackage{todonotes}
\usepackage{csvsimple}
\usepackage{multirow}
\usepackage{multicol}
\usepackage[hidelinks, breaklinks]{hyperref}

\usepackage[T1]{fontenc}
\usepackage{float}
\usepackage{verbatim}
\usepackage{textcomp}
\usepackage{breakurl}

\usepackage{algorithm}
\usepackage{algorithmicx}
\usepackage{algpseudocode}

\newcommand{\CC}{C\nolinebreak\hspace{-.05em}\raisebox{.4ex}{\tiny\bf +}\nolinebreak\hspace{-.10em}\raisebox{.4ex}{\tiny\bf +}}
\def\CC{{C\nolinebreak[4]\hspace{-.05em}\raisebox{.4ex}{\tiny\bf ++}}}

\begin{document}
\frontmatter
\pagestyle{headings}
\mainmatter              % start of the contributions
\title{Atari games and Intel processors}
\titlerunning{Atari Games and Intel processors}  % abbreviated title (for running head)

\author{Robert Adamski, Tomasz Grel, Maciej Klimek and Henryk Michalewski}
\authorrunning{R. Adamski et al.} % abbreviated author list (for running head)
\institute{Intel, deepsense.io, University of Warsaw\\
\email{Robert.Adamski@intel.com, T.Grel@deepsense.io, M.Klimek@deepsense.io, H.Michalewski@mimuw.edu.pl}}

\maketitle
\begin{abstract}
The asynchronous
nature of the state-of-the-art reinforcement learning algorithms such as the Asynchronous Advantage Actor-Critic algorithm, makes them exceptionally suitable for CPU computations. 
However, given the fact that deep reinforcement learning often deals with interpreting visual information, a large part of the train and inference time is spent performing convolutions.

%In this case correct utilization of wide Single Instruction Multiple Data paradigm
%and parallel programming techniques seem to be crucial for achieving optimal performance. 
In this work we present our results on learning strategies in Atari games using a Convolutional Neural Network, the Math Kernel Library and TensorFlow 0.11rc0 machine learning framework. We also analyze effects of asynchronous computations on the convergence of reinforcement learning algorithms.
\end{abstract}

\keywords{reinforcement learning, deep learning, Atari games, asynchronous computations}

\section{Introduction}
In this work we approach the problem of learning strategies in Atari games from the hardware architecture perspective. %complementing a work done in \cite{GA3C} with information on from the CPU perspective.  
%technical and low-level perspective. 
We use a variation of the statistical model developed in \cite{DeepMindA3CPaper,AtariNature}. %However, 
Using the provided code\footnote{\url{https://github.com/deepsense-io/BA3C-CPU}} our experiments are easy % not only to satisfy sheer engineering curiosity, but also 
%  in order to make sure that everyone can 
to re-create and we encourage the reader to draw his own conclusions 
%run on its own computer and draw own conclusions 
about how CPUs perform in the context of Atari games. Following \cite{gym-whitepaper,DeepMindA3CPaper,AtariNature} we treat Atari games as a key benchmark problem for modern reinforcement learning.

We use a statistical model consisting of approximately one million %\todo{Tomek: checked -- the model has around 950k parameters} 
floating point numbers which are iteratively updated using a gradient descent algorithm described in \cite{adam}. %\todo{Cite the right algorithm}. 
% In principle the update can be done according to a favorite reinforcement learning algorithm. %\todo{Cite here Sutton Barto, a specific place}. 

At first glance a training of such model appears as a relatively straightforward task: a screen from the simulator is fed into the statistical model which decides which button must be pressed; over an episode of a game we estimate how the agent performs and calculate the loss accordingly and update the model so that the loss is reduced. 

In practice filling details of the above scenario is quite challenging. In this work we accept a number of technical solutions presented in \cite{DeepMindA3CPaper}. %Since the article \cite{DeepMindA3CPaper} does not provide too many technical details, we fill these details on our own.  
Our work also follows closely research done in \cite{GA3C}, where a batch version of \cite{DeepMindA3CPaper} is analyzed. We describe our algorithmic decisions in considerable detail in Section \ref{subsection:ba3c_details}.

We obtained state-of-the-art results in all tested games (see Section \ref{section:results}) and in the process of obtaining them we detected certain interesting issues described in Sections \ref{subsection:effects}, \ref{subsection:batch} related to batch sizes, learning rates and the asynchronous learning algorithm we use in this paper. The issues are illustrated by Figures \ref{delay_fig_1} and \ref{delay_fig_2}. Apparently our algorithm relies on timely emptying of queues. If queues are growing, then updates are delayed and learning performance degenerates up to the point where the trained agent goes back to an essentially a random behavior. This in turn implies certain ``preferred'' sizes of batches as illustrated by Figure  \ref{figure:overall}. Those batch sizes in turn imply ``preferred'' learning rates also visible in Figure  \ref{figure:overall}.

Our contribution can be considered as a snapshot of the CPU performance in the domain of reinforcement learning illustrating engineering opportunities and obstacles one can encounter relying solely on a CPU hardware. We also contributed an integration of Google's machine learning framework TensorFlow 0.11rc0 with Intel's Math Kernel Library (MKL). Details of the integration are described in Section \ref{section:changes} and benchmarks comparing behavior of the out-of-the-box TensorFlow 0.11rc0 with our modified version are included in Section \ref{subsection:benchmark}.  Section \ref{section:hardware} contains a description of our hardware. Let us underline that we relied on a completely standard Intel servers. Section \ref{section:mkl} contains a brief characteristic of the MKL library and its currently available deep learning functionalities.

The learning times on our hardware  described in Section \ref{section:hardware} are very competitive (see Figure \ref{plot:score_vs_time}) and in a future work we are planning to 
%generalize results of \cite{allinea} and 
bring it down to minutes using sufficiently large CPU clusters. 
%on a modern server is contained within  

%strategy which maximizes agents' perception of the 
% No similar open-sourced CPU-centric experiments at the moment when the research started. The problem with queues. 

\subsection{Related tools}

This work would be impossible without a number of custom machine learning and reinforcement learning engineering tools. Our work is based on
\begin{itemize}
\item OpenAI Gym \cite{gym-whitepaper}, an open source machine learning platform allowing a very easy access to a rich library of games, including Atari games,
\item Google's TensorFlow 0.11rc0, an open source machine learning framework \cite{tensorflow2015-whitepaper} allowing for streamlined integration of various neural networks primitives (layers) implemented elsewhere, 
\item Tensorpack, an open source library \cite{Tensorpack} implementing a very efficient reinforcement learning algorithm, 
\item Intel's Math Kernel Library 2017 (MKL) \cite{intel-mkl_dnn}, a freely available library which implemented neural networks primitives (layers) and overall speeds up matrix and in particular deep learning computations on Intel's processors. 
\end{itemize}

\subsection{Related work}

\paragraph{Relation to \cite{DeepMindA3CPaper}.}
%The key reference for our work is \cite{DeepMindA3CPaper} and we first describe how our paper is related to 

\noindent
{\it Decisions in which we follow \cite{DeepMindA3CPaper}.}
One of the key decisions is to run many independent agents in separate environments in an asynchronous way. In the training process in every environment we play an episode of 2000  
steps (the number may be smaller if the agent dies).
An input to the statistical model consists of 4 subsequent screens in the RGB format. An output of the statistical model is one of 18 possible moves of the controller. 
Over each episode the agent generates certain reward. The reward allows us to estimate how good were decisions made for every screen appearing during the episode. At every step an impact of the reward on decisions made earlier in the episode is discounted by a factor $\gamma$ ($0<\gamma\leq 1$).

Having computed rewards for a given episode we can update weights of the model according to rewards --- this is done through gradient updates which are applied directly to the statistical model weights. The updates are scaled by a learning rate $\lambda$. Authors of \cite{DeepMindA3CPaper} reported good CPU performance and this encouraged the experiment described in this paper. 
 
\noindent
{\it Decisions left to readers of \cite{DeepMindA3CPaper}.}
The key missing detail are all technical decisions related to communication between processes. % Overall, we are not aware of any publicly available high-performing implementation which would follow details of \cite{DeepMindA3CPaper}. %which seems to indicate that various  

\paragraph{Relation to \cite{GA3C} and \cite{allinea}.}
Since the publication of \cite{AtariNature} a significant number of new results was obtained in the domain of Atari games, however to the best of our knowledge only the works \cite{GA3C} and \cite{allinea} were focused on the hardware performance. In \cite{GA3C} authors modify the approach from \cite{DeepMindA3CPaper} so it fits better into the GPU multi-core infrastructure. In this work we show that a similar modification can be also quite helpful for the CPU performance. This work can be considered a CPU variant of \cite{GA3C}. In \cite{allinea} a significant speedup of the A3C algorithm was obtained using large CPU clusters. However, it is unclear if the method scales beyond the game of Pong. Also the announcement  \cite{allinea} does not contain neither technical details or implementation.  % very little is said about the correctness of the resulting models.  %there is no single article which would focus on the analysis of learning performance from the hardware perspective.

\paragraph{Relation to \cite{intel_tensorflow}.} The fork of TensorFlow announced in \cite{intel_tensorflow} will offer a much deeper integration of TensorFlow and Intel's Math Kernel Library (MKL). In particular it should resolve the dimensionality issue mentioned in Section \ref{subsection:possible_imp}. However, at the moment of writing of this paper we had to do the integration of these tools on our own, because the fork mentioned in  \cite{intel_tensorflow} was not ready for our experiments. 

\paragraph{Other references.}
The work \cite{AtariNature} approaches the problem of learning a strategy in Atari games through approximation of the $Q$-function, that is implicitly it learns a synthesized values of every move of a player in a given situation on the screen. We did not consider this method, because of overall weaker results and much longer training times comparing to the asynchronous methods in \cite{DeepMindA3CPaper}.

The DeepBench \cite{DeepBench}, the FALCON Library \cite{Falcon} and the study \cite{intel_mkl_alexnet} compare a performance of CPU and GPU on neural network primitives (single convolutional and dense layers) as well as on a supervised classification problem. Our article can be considered a reinforcement learning variant of these works. % to the realm of . %  we see GEMM computations and convolution operations on Xeon Phi CPU at level comparable to GPU M40, and in some cases to Titan architecture. That shows great potential for DNN  optimization efforts on CPU.  %\todo[inline]{This should be just a regular reference (github link would end up in the bibliography).} 

A recently published work \cite{openaiES} shows a very promising CPU-only results for agent training tasks. The learning algorithm proposed  in \cite{openaiES} is a novel approach with yet untested stability properties. Our work focuses on a more established family of algorithms with better understood theoretical properties and applicability tested on a broader class of domains.

For a broad introduction to reinforcement learning we refer the reader to \cite{BartoSutton}. For a historical background on Atari games we refer to \cite{AtariNature}.
%This is an interesting, but rather a slow method with learning times counting in days.  

%Underline that this is A3C algorithm for simple people who want to repeat it on their own hardware. 

%\subsection{Introduction to reinforcement learning}
%Underline relative simplicity of the basic algorithm comparing to value-based algorithms. 

%In the area of Reinforcement Learning we deal with one or more agents\ldots

%\subsection{Historical results about Atari games}

%Here we should also mention some old stuff about MCTS, not just neural nets. 

\section{The Batch Asynchronous Advantage Actor Critic Algorithm (BA3C)}
The Advantage Actor Critic algorithm (A2C) is a reinforcement learning algorithm combining positive aspects of both policy-based and value function based approaches to reinforcement learning. 
The results reported recently by Mnih et al. in \cite{DeepMindA3CPaper} provide strong arguments for using its asynchronous version (A3C). After testing several implementations of this algorithm we found that a high quality open source implementation of this algorithm is provided in the TensorPack (TP) framework \cite{Tensorpack}. However, the differences between this variant, which resembles an algorithm introduced in \cite{GA3C}, and the one described originally in \cite{DeepMindA3CPaper} are significant enough to justify a new name. Therefore we will refer to this implementation as the Batch Asynchronous Advantage Actor Critic (BA3C) algorithm.\footnote{In \cite{GA3C} is proposed a different name GA3C derived from ``hybrid CPU/GPU implementation of the A3C algorithm''. This seems a bit inconvenient, because it suggests a particular link between the batch algorithm and the GPU hardware; in this work we obtain good results for a similar algorithm running only on CPU. } % A short explanation of the differences between it and the standard A3C follows. 
%We are not aware of a high-performing open-source implementations of the A3C algorithm which would not use 

\subsection{Asynchronous reinforcement learning algorithms}

Asynchronous reinforcement learning procedures are designed to use multiple concurrent environments to speed up the training process. This leaves an issue how the model or models are stored and synchronized between environments. We discuss some possible options in \ref{subsection:ba3c_details}, including description of our own decisions. % Traditionally, one ``local'' model is used for each training environment and the updates accumulated by these models are then combined to update the global model, to which the local models are synchronized.

Apart from obvious speedups resulting from utilizing concurrency, this approach has also some statistical consequences. Usually in one environment the subsequent states are highly correlated. This can have some adverse effects on the training process. However, when using multiple environments simultaneously, the states in each environment are likely to be significantly different, thus decorrelating the training points and enabling the algorithm to converge even faster.

\subsection{BA3C -- details of the implementation}
\label{subsection:ba3c_details}

The batch variant of the A3C algorithm was designed to better utilize massively parallel hardware by batching data points. % in a manner similar to what one could use in supervised learning. 
Multiple environments are still used, but there's only one instance of the model. This forces the extensive use of threading and message queues to decouple the part of the algorithm that generates the data from the one responsible for updates of the model. % What follows is an explanation of the computational structure of the BA3C algorithm.
%In the most simple case, 
In a simple case of only one environment the BA3C algorithm consists of the steps described in algorithm \ref{alg:sync_rl}.

\begin{algorithm}
\caption{Basic synchronous Reinforcement Learning scheme}
\label{alg:sync_rl}
\begin{algorithmic}[1]
\State Randomly initialize the model.
\State Initialize the environment.
\Repeat
  \State Play $n$ episodes by using the current model
     to choose optimal actions.
  \State Memorize obtained states and rewards.
  \State Use the generated data points to train and update the model.
\Until results are satisfactory.
\end{algorithmic}
\end{algorithm}

\begin{figure}[H]
\centering
\includegraphics[width=0.7\textwidth]{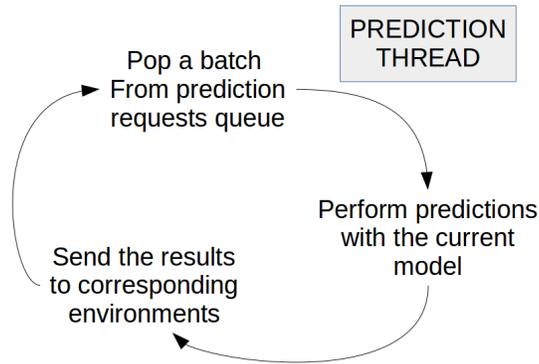}
\caption{Main loop of the prediction thread, which is responsible for evaluating the state of the environment and choosing the best action based on the current policy model.}
\label{fig:prediction_thread}
\end{figure}

\begin{figure}[H]
\centering
\includegraphics[width=0.7\textwidth]{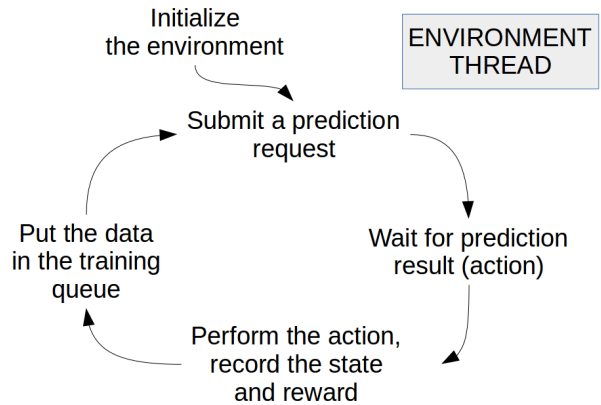}
\caption{Main loop of a single environment thread. Usually multiple environment threads will be working in parallel in order to generate the training data faster.}
\label{fig:environment_thread}
\end{figure}

When using multiple environments one can follow a similar approach - each environment could simply use the global model to predict the optimal action given its current state. Let us notice that
the model always performs prediction on just a single data point from a single environment (i.e.: a single state vector of the environment). Obviously, this is far from optimal in terms of processing speed. Also accessing the shared model from different environments will quickly become a bottleneck. The two most popular approaches for solving this problem are:
\begin{itemize}
\item Maintaining several local copies of the model (one for each environment) and synchronizing them with a global model. This approach is used and extensively described in \cite{DeepMindA3CPaper,Gorilla,Hogwild} and we  refer to it as A3C. 
%\todo{CHECKED! this is only mentioned in the A3C paper, but described in details in the earlier gorilla paper by deepmind}
\item Using a single model and batching the predictions from multiple environments together (the ``batch'' variant, BA3C). This is much more suitable for use on massively parallel hardware \cite{GA3C}.
%\todo{directly from nvidia paper: When using a GPU, the mix of
%small DNN architectures and small batch sizes can severely underutilize computational resources}
\end{itemize}
The batch variant requires using the following queues for storing data:
\begin{description}
\item[Training queue] -- stores the data points generated by the environments; the data points are used in training. See Figure \ref{fig:training_thread}.
\begin{figure}[H]
\centering
\includegraphics[width=0.6\textwidth]{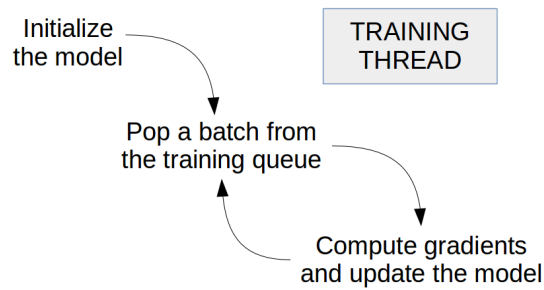}
\caption{Activities performed by the training thread. Please note that popping the data from the training queue may involve waiting until the queue has enough elements in it.}
\label{fig:training_thread}
\end{figure}
\item[Prediction requests queue] 
-- stores the prediction requests made by the environments; the predictions are made according to the current weights stored in the model. See Figure \ref{fig:prediction_thread}.
\item[Prediction results queue] -- stores the results of the predictions made by the model; the predictions are later used by the environments for choosing actions. See Figure \ref{fig:environment_thread}.
\end{description}

\subsubsection{Hyperparameters}
In Table \ref{table:hyper} we list the most important hyperparameters of the algorithm is presented.

\begin{table}
\centering
\caption{Description of the hyperparameters of the algorithm.}
\label{table:hyperparameters}
{\small 
\begin{tabular}{l c p{5cm}} 
parameter & default value & description \\\hline
learning rate & 0.001 & step size for the optimization algorithm\\
batch size & 128 & number of training examples in a training batch \\
frame history & 4 & the number of consecutive frames to take into consideration while evaluating the current state of the game \\
local time max & 5 & number of consecutive data points to memorize before concluding the episode with a reward estimate based on the output of the value network \\
image size & (84,84) & the size to which to rescale the original input into. This is done mainly because working on the original images is very expensive. \\
gamma & 0.99 & the discount factor \\
\end{tabular}}
\label{table:hyper}
\end{table}

\subsubsection{ConvNet architecture}
We made rather minor changes to the original TensorPack ConvNet. The main focus of the changes was to better utilize the MKL convolution primitives to enhance the performance. The architecture is presented in the diagram below.

\begin{figure}[H]
\centering
\includegraphics[width=0.5\textwidth]{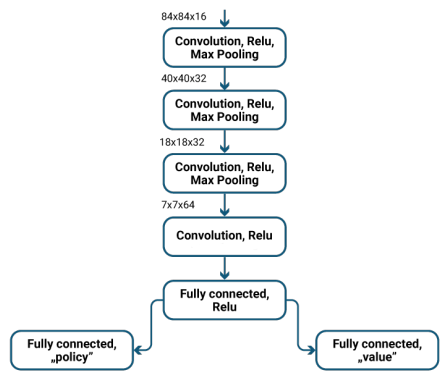}
\caption{The structure of the Convolutional Neural Network used for processing the input images}
\label{conv_net_diagram}
\end{figure}

\subsection{Effects of asynchronism on convergence}
\label{subsection:effects}

Training and prediction part of the above described algorithm work in separate threads and there's a possibility that one of those parts will work faster than the other (in terms of data points processed per unit time). This is rarely an issue when the training thread is faster -- in this case it'll simply find out that the training queue is empty and wait for a batch of training data to be generated. This is inefficient since the hardware is not fully utilized when the train thread is waiting for data, but it should not impact the correctness of the algorithm.

A much more interesting case arises when data points are generated faster than can be consumed by the training thread. If we're using default first-in-first-out training queue and this queue is not empty, then there's some delay between the batch of data being generated by the prediction thread and it being used for training. It turns out that if this delay is large enough it will have detrimental effect on the convergence of the algorithm.

When there's a significant delay between the generation of a batch and training on it, the training will be performed using a data point generated by an older model. That is because when the batch of data was ``waiting'' in the training queue, other batches were used for training and the model was updated. The number of such updates is equal to the size of the queue at the time when this batch was generated. Therefore the updates are performed using out-of-date training data which may have little to do with the current policy maintained by the current model.

Of course when this delay is small and the learning rate is moderate the current policy is almost equal to the ``old'' one used for generating the training batch and the training process will converge. In other cases one should have means of constraining the delay to force correct behavior.

The solution is to restrict the size of the training queue. This way, when the training thread is generating too many training batches it will at some point reach the full capacity of the queue and will be forced to wait until some batch is popped. Usually the the size of the training queue is set to ensure that the training can take place smoothly. What we found out, however, is that setting the queue capacity to extremely small values (i.e., less than five), has little if any impact on the overall training speed.

\subsubsection{Impact of delay on convergence -- experiments}

This section describes a series of experiments we've carried out in order to establish how big a delay in the pipeline has to be to negatively impact the convergence. The setup involved inserting a fixed size first-in-first-out buffer between the prediction and training parts of the algorithm. This buffer's task was to ensure a predefined delay in the algorithm was present. With this modification we were able to conduct a series of experiments for different sizes of this buffer (delays). The results are shown below. 

\begin{center}
\begin{figure}[H]
\centering
\includegraphics[width=0.6\textwidth, bb=0 0 576 432]{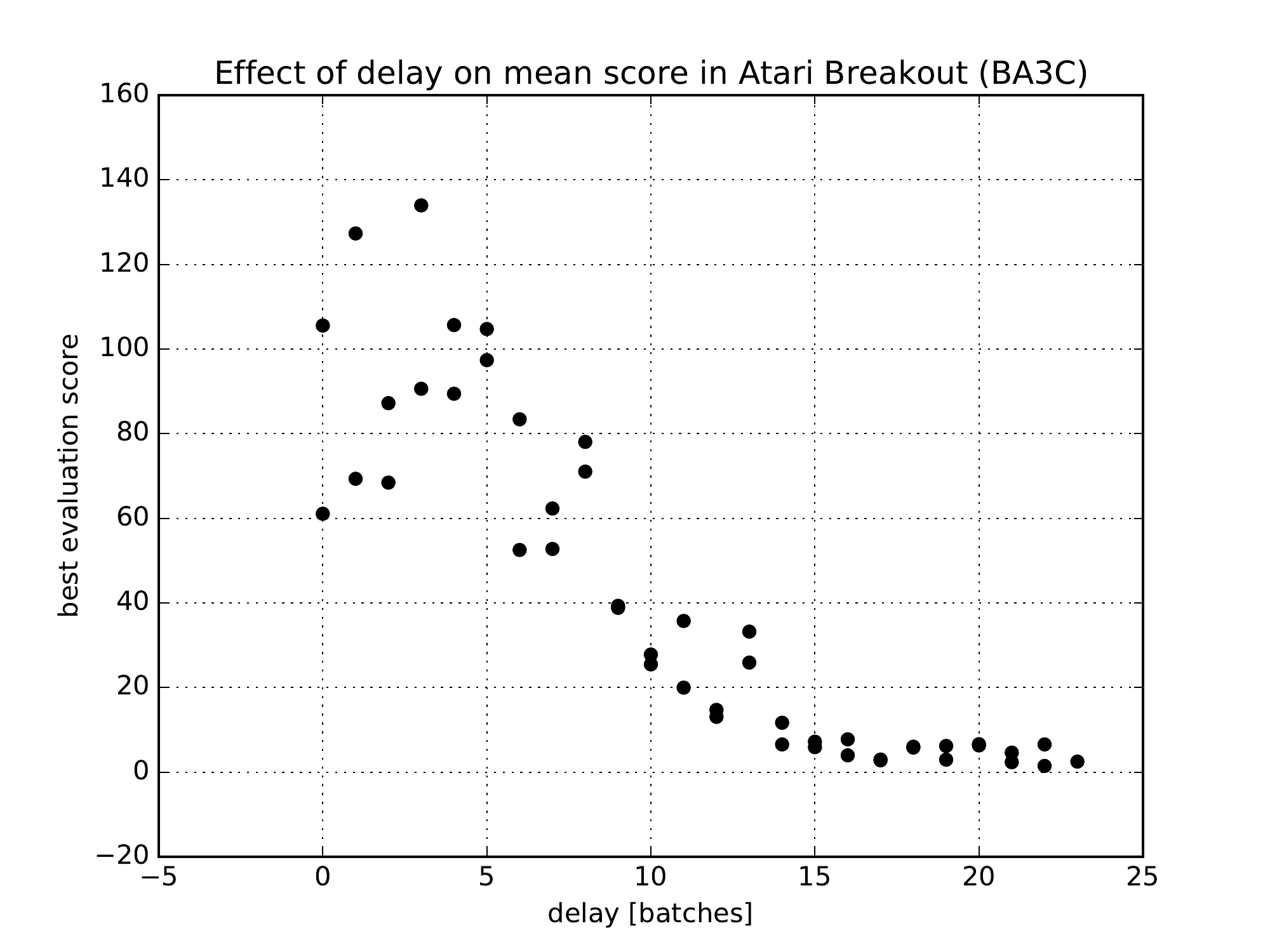}
\caption{Best evaluation results for experiments with different artificial delays introduced into the pipeline. For this experiment the default batch size of 128 was used. It seems that even very small delays have a negative impact, while a delay of more than 10 batches (i.e.: $10 \cdot 128=1280$ data points when using the default batch size of 128) is enough to totally prevent the algorithm from convergence.}
\label{delay_fig_1}
\end{figure}
\end{center}
\begin{figure}[H]
\centering
\includegraphics[width=0.6\textwidth, bb=0 0 576 432]{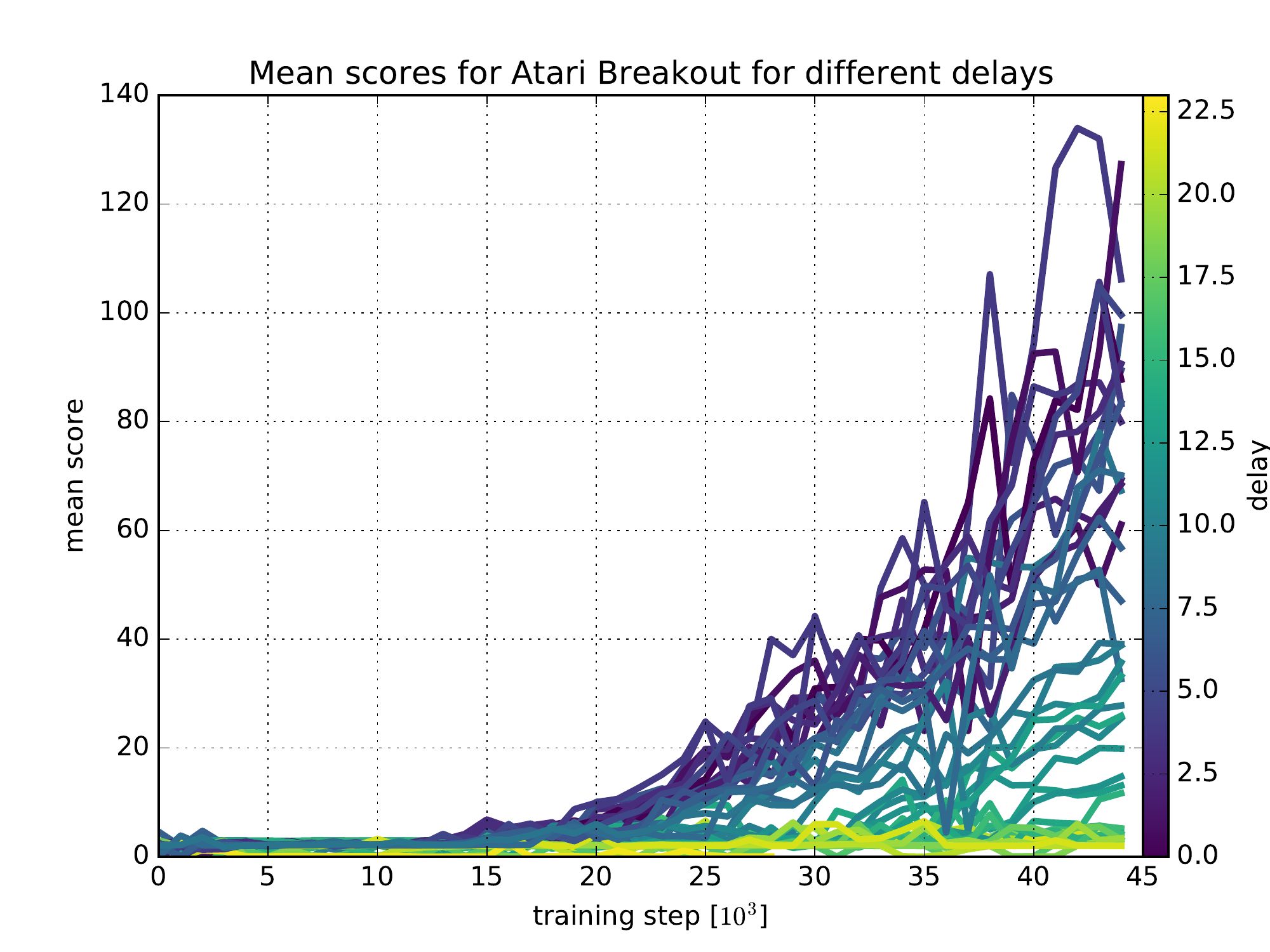}
\caption{Mean scores for Atari breakout for different delays. The plot shows the course of learning for the artificial delays in the pipeline varying between 0 and 23, the brighter the line, the more delay was introduced. It is visible that a delay greater than 10 can prevent the algorithm from successful convergence.}
\label{delay_fig_2}
\end{figure}

Based on our results presented in the figures \ref{delay_fig_1} and \ref{delay_fig_2} we can conclude that even small delays have significant impact on the results and delays of more than 10 batches (1280 data points) effectively prevented the BA3C from converging. Therefore when designing an asynchronous RL algorithm it might be a good idea to try to streamline the pipeline as much as possible by making the queues as small as possible. This should not have significant effects on processing speed and can significantly improve obtained results.

\section{Specification of involved hardware}
\label{section:hardware}

\subsection{Intel Xeon\textsuperscript{\textregistered}\ (Broadwell)}
We used Intel
Xeon\textsuperscript{\textregistered}\ E5 2600 v4 processors
%referred later in this article as ‘Xeon’ was used 
to perform benchmarks tests of convolutions.
Xeon Broadwell is based on processor microarchitecture known as a ``tick'' \cite{intel-Broadwell} -- a die shrink of an existing architecture, rather than a new architecture.  In that sense, Broadwell is basically a Haswell made on Intel's 14nm second generation tri-gate transistor process with few improvements to the micro-architecture. Important changes are: up to 22 cores per CPU; support for DDR4 memory up to 2400 MHz; faster floating point instruction performance; improved performance on large data sets.
Results reported here are obtained 	on a system with two  Intel Xeon\textsuperscript{\textregistered}\ Processor E5 2689  (3.10 GHz, 10  core) with 128 GB of DDR4 2400MHz RAM, Intel Compute Module S2600TP and Intel Server Chassis H2312XXLR2. 
The system was running Ubuntu 16.04 LTS operating system. The code was compiled with 
GCC 5.4.0 and linked against the Intel MKL 2017 library (build date 20160802). 

\subsection{ Intel\ Xeon\textsuperscript{\textregistered}\ (Haswell)}
 Intel Xeon\textsuperscript{\textregistered}\ E5 2600 v3 Processor, was used as base for series of experiments to test hyperparameters of our algorithm. %, results in chapter 8. 
Haswell brings, along with new microarchitecture, important features like AVX2.
We used 
%TOP500 
the Prometheus cluster with a peak performance of 2.4 PFlops located at the Academic Computer Center Cyfronet AGH as our testbed platform. 
Prometheus  consists of more than 2,200 servers, accompanied by 279 TB RAM in total, and by two storage file systems of 10 PB total capacity and 180 GB/s access speed.
Experiments were performed in single-node mode, each node consisting of two Intel Xeon\textsuperscript{\textregistered}\  E5-2680v3 processors with 24 cores at 2.5GHz with 128GB of RAM, with peak performance of 1.07 TFlops. 

Xeon Haswell CPU allows effective computations of CNN algorithms, and convolutions in particular, by taking advantage of SIMD (single instruction, multiple data) instructions via vectorization and of multiple compute cores via threading. Vectorization is extremely important as these processors operate on vectors of data up to 256 bits long (8 single-precision numbers) and can perform up to two multiply and add (Fused Multiply Add, or FMA) operations per cycle.
Processors support Intel Advanced Vector Extensions 2.0 (AVX2) vector-instruction sets
which provide:
(1) 256-bit floating-point arithmetic primitives, 
(2) Enhancements for flexible SIMD data movements.
These architecture-specific advantages have been implemented in the Math Kernel Library (MKL) and used in deep learning framework Caffe  \cite{intel-dubey}, \cite{intel-caffe_optimized} resulting in improved convolutions performance.

\section{The MKL library}
\label{section:mkl}

The Intel Math Kernel Library (Intel MKL) 2017 introduces a set of Deep Neural Networks (DNN) \cite{intel-mkl_dnn} primitives for DNN applications optimized for the Intel architecture. %The implementation of DNN functions includes a set of primitives necessary to accelerate popular image recognition topologies, such as AlexNet, Visual Geometry Group (VGG), GoogleNet, and Residual Networks (ResNet).
The primitives implement forward and backward passes for the following operations:
% \begin{itemize}
%\item 
(1) Convolution: direct batched convolution,
%\item 
(2) Inner product,
% \item 
(3) Pooling: maximum, minimum, and average,
%\item 
(4) Normalization: local response normalization across channels and batch normalization,
% \item 
(5) Activation: rectified linear neuron activation (ReLU), 
% \item 
(6) Data manipulation: multi-dimensional transposition (conversion), split, concatenation, sum, and scale.
% \end{itemize}
Intel MKL DNN primitives implement a plain C application programming interface (API) that can be used in the existing C/\CC\ DNN frameworks, as well as in custom DNN applications.

\section{Changes in TensorFlow 0.11rc0}
\label{section:changes}

\subsection{Motivation}
Preliminary benchmarks showed that the vast majority of computation time during training is spent performing convolutions. On CPU the single most expensive operation was the backward pass with respect to the convolution's kernels, especially in the first layers working on the largest inputs. Therefore significant increases in performance had to be achieved by optimizing the convolution operation.

We considered the following approaches to this problem:
\begin{description}
\item [Tuning the current implementation of convolutions] -- TensorFlow (TF) uses the Eigen \cite{eigenweb} library as a backend for performing matrix operations on CPU. Therefore this approach would require performing changes in the code of this library. The matrix multiplication procedures used inside Eigen have multiple hyperparameters that determine the way in which the work is divided between the threads. Also, some rather strong assumptions about the configuration of the machine (e.g., its cache size) are made. %It is also unlikely anyone benchmarked this implementation on a 20-core Broadwell machine and even more unlikely that it has been tested on a Knight's Landing (KNL) processor, which was officially released only a few months earlier. 
This certainly leaves space for improvements, especially when optimizing for a very specific use-case and hardware.

\item [Providing alternative implementation of convolutions] -- The MKL library provides deep neural network operations optimized for the Intel architectures. %Therefore it is far more likely that the procedures in MKL were benchmarked against our hardware configuration. 
Some tests of convolutions on a comparable hardware
%the KNL chip 
had already been performed by Baidu \cite{DeepBench} and showed promising results. This also had the added benefit of leaving the original implementation unchanged thus making it possible for the user to decide which implementation (the default or the optimized one) to use.
\end{description}

We decided to employ the second approach that involved using the MKL convolution. A similar decision was taken also in the development of the Intel-focused fork of TensorFlow \cite{intel_tensorflow}.

\subsection {Implementation}

TensorFlow provides a well documented mechanism for adding user-defined operations in \CC, which makes it possible to load additional operations as shared objects. However, maintaining a build for a separate binary would make it harder to use some internal TF's utilities and sharing code with the original convolution operation. Therefore we decided to fork the entire framework and provide the additional operations.

Another TF's feature called 'labels' made it very simple to provide several different implementations of the same operation in \CC\ and choose between them from the python layer by specifying a 'label map'. This proved especially helpful while testing and benchmarking our implementation since we could quickly compare it to the original implementation.

The implementation consisted of linking against the MKL library and providing the three additional operations:
%\begin{itemize}
%\item 
(1) MKL convolution forward pass,
%\item 
(2) MKL convolution backpropagation w.r.t. the input feature map,
%\item 
(3) MKL convolution backpropagation w.r.t. the kernels.
%\end{itemize}

The code of these operations formed a glue layer between the TF's and MKL's programming interfaces. The computations were performed inside highly optimized MKL primitives.
\subsection{Benchmark results}
\renewcommand{\arraystretch}{1.5}
\begin{table}[H]
\centering
\caption{Forward convolution times [ms]. Notice that the MKL TF times are consistently smaller than the standard TF times. Data layout conversion times are not included in these measurements.}
\label{forward_table}
\begin{tabular}{|l|l|l|l|l|l|}
\hline
\multirow{2}{*}{input size} & \multirow{2}{*}{kernel size} & \multicolumn{2}{l|}{MKL TF}                           & \multicolumn{2}{l|}{TF}          \\ \cline{3-6} 
                            &                              & \multicolumn{1}{l|}{Phi} & \multicolumn{1}{l|}{Xeon} & \multicolumn{1}{l|}{Phi} & Xeon  \\ \hline
128,84,84,16                & 16,32,5,5                    & 10.03                    & 23.61                     & 90.11                    & 99.74 \\ \cline{1-2}
128,40,40,32                & 32,32,5,5                    & 4.58                     & 8.76                      & 43.83                    & 33.61 \\ \cline{1-2}
128,18,18,32                & 32,64,5,5                    & 1.61                     & 2.71                      & 17.20                    & 10.22 \\ \cline{1-2}
128,7,7,64                  & 64,64,3,3                    & 0.88                     & 0.38                      & 3.50                     & 0.79  \\ \hline
\end{tabular}
\end{table}
\label{subsection:benchmark}
Multiple benchmarks were conducted in order to assess the performance of our implementation. They are focused on a specific 4-layer ConvNet architecture used for processing the Atari input images. The results are shown below.

Tables \ref{forward_table}, \ref{backward_data_table}  and \ref{backward_filter_table} show the benchmark results for the TensorFlow modified to use MKL and standard TensorFlow. Measurements consist of the times of performing convolutions with specific parameters (input and filter sizes) for Xeon\textsuperscript{\textregistered}\ and Xeon Phi\textsuperscript{\textregistered}\ CPUs. The same convolution parameters were used in the convolutional network used in the atari games experiments.

The results show that the MKL convolutions can be substantially faster than the ones implemented in TensorFlow. For some operations a speed-up of more than 10 times was achieved. The results agree with the ones reported in \cite{DeepBench}. It is also worth noticing that most of the time is spent in the first layer which is responsible for processing the largest images.

\begin{table}[H]
\centering
\caption{Backward data convolution times [ms]. TensorFlow times for the first layer are not listed since computing the gradient w.r.t the input of the model is unnecessary.}
\label{backward_data_table}
\begin{tabular}{|l|l|l|l|l|l|}
\hline
\multirow{2}{*}{input size} & \multirow{2}{*}{kernel size} & \multicolumn{2}{l|}{MKL TF}                           & \multicolumn{2}{l|}{TF}           \\ \cline{3-6} 
                            &                              & \multicolumn{1}{l|}{Phi} & \multicolumn{1}{l|}{Xeon} & \multicolumn{1}{l|}{Phi} & Xeon   \\ \hline
128,84,84,16                & 16,32,5,5                    & N/A                      & N/A                       & N/A                      & N/A    \\ \cline{1-2}
128,40,40,32                & 32,32,5,5                    & 11.17                    & 16.99                     & 468.82                   & 112.77 \\ \cline{1-2}
128,18,18,32                & 32,64,5,5                    & 4.38                     & 4.55                      & 50.09                    & 9.74   \\ \cline{1-2}
128,7,7,64                  & 64,64,3,3                    & 2.14                     & 0.77                      & 4.41                     & 1.22   \\ \hline
\end{tabular}
\end{table}

\begin{table}[H]
\centering
\caption{Backward filter convolution times [ms]. Please note very long time spent in the first layer by the standard TensorFlow convolution. It was possible to reduce it more than 10 times by using our implementation}
\label{backward_filter_table}
\begin{tabular}{|l|l|l|l|l|l|}
\hline
\multirow{2}{*}{input size} & \multirow{2}{*}{kernel size} & \multicolumn{2}{l|}{MKL TF}                           & \multicolumn{2}{l|}{TF}                              \\ \cline{3-6} 
                            &                             & \multicolumn{1}{l|}{Phi} & \multicolumn{1}{l|}{Xeon} & \multicolumn{1}{l|}{Phi} & \multicolumn{1}{l|}{Xeon} \\ \hline
128,84,84,16                & 16,32,5,5                   & 8.97                     & 29.63                     & 1,236.98                 & 368.18                    \\ \cline{1-2}
128,40,40,32                & 32,32,5,5                   & 6.33                     & 19.55                     & 343.73                   & 114.72                    \\ \cline{1-2}
128,18,18,32                & 32,64,5,5                   & 2.52                     & 6.07                      & 36.74                    & 28.82                     \\ \cline{1-2}
128,7,7,64                  & 64,64,3,3                   & 2.31                     & 3.18                      & 7.38                     & 5.57                      \\ \hline
\end{tabular}
\end{table}

\subsection{Possible improvements}
\label{subsection:possible_imp}

%TODO : add a citation in the first sentence?
%In CPU computations on multicore,  SIMD architectures 
The data layout can have a tremendous impact on performance of low-level array operations. In turn, efficiency of these operations is critical for performance of higher-level machine learning algorithms. 
%in a way which benefits from multicore and SIMD architectures. 

TensorFlow and MKL have radically different philosophies of storing visual data. TensorFlow uses mostly its default ``NHWC'' format, in which pixels with the same spatial location but different channel indices are placed close to each other in memory. Some operations also provide the ``NCHW'' format widely used by other deep learning frameworks such as Caffe \cite{caffe}. On the other hand MKL does not have a predefined default format, rather it is designed to easily connect MKL layers to one another. In particular, the same operation can require different data layouts depending on the sizes of its input (e.g. the number of input channels). This is supposed to ensure that the number of intermediate ``conversions'' or ``transpositions'' in the pipeline is minimal, while at the same time letting each operation use its preferred data layout. 

It is important to note that our implementation provided an alternative ``MKL'' implementation only for the convolution. We did not provide similar alternatives for max pooling, ReLU etc. This forced us to repeatedly convert the data between the TF's NHWC format and the formats required by the MKL convolution. Obviously this is not an optimal approach, however, implementing it optimally would most probably require significant changes in the very heart of the framework -- its compiler. This task was beyond the scope of the project, but it's certainly feasible and with enough effort our implementation's performance could be even further improved.
The times necessary to perform data conversions are provided in the Table \ref{conversions_table}.

\begin{table}[H]
\centering
\caption{Data layout conversion times [ms].}
\label{conversions_table}
\begin{tabular}{|l|l|l|l|l|l|l|l|}
\hline
\multirow{2}{*}{input size} & \multirow{2}{*}{kernel size} & \multicolumn{2}{l|}{Forward}      & \multicolumn{2}{l|}{BWD Filter} & \multicolumn{2}{l|}{BWD data} \\ \cline{3-8} 
                            &                              & Phi& Xeon & Phi   & Xeon & Phi  & Xeon \\ \hline
128,84,84,16                & 16,32,5,5                    & 37.44             & 12.20         & 14.55               & 11.70          & N/A                & N/A           \\
128,40,40,32                & 32,32,5,5                    & 3.30              & 2.92          & 5.32                & 4.34           & 6.18               & 4.14          \\
128,18,18,32                & 32,64,5,5                    & 2.31              & 0.58          & 4.32                & 0.62           & 5.89               & 0.68          \\
128,7,7,64                  & 64,64,3,3                    & 1.96              & 0.15          & 11.48               & 0.56           & 2.59               & 0.24          \\ \hline
\end{tabular}
\end{table}

%\subsection{The MKL library}
\section{Results}
\label{section:results}

\subsection{Game scores and overall training time}
\begin{figure}[H]
%\centering
\hspace*{-1cm}
\includegraphics[width=1.2\textwidth, bb=0 0 481 155]{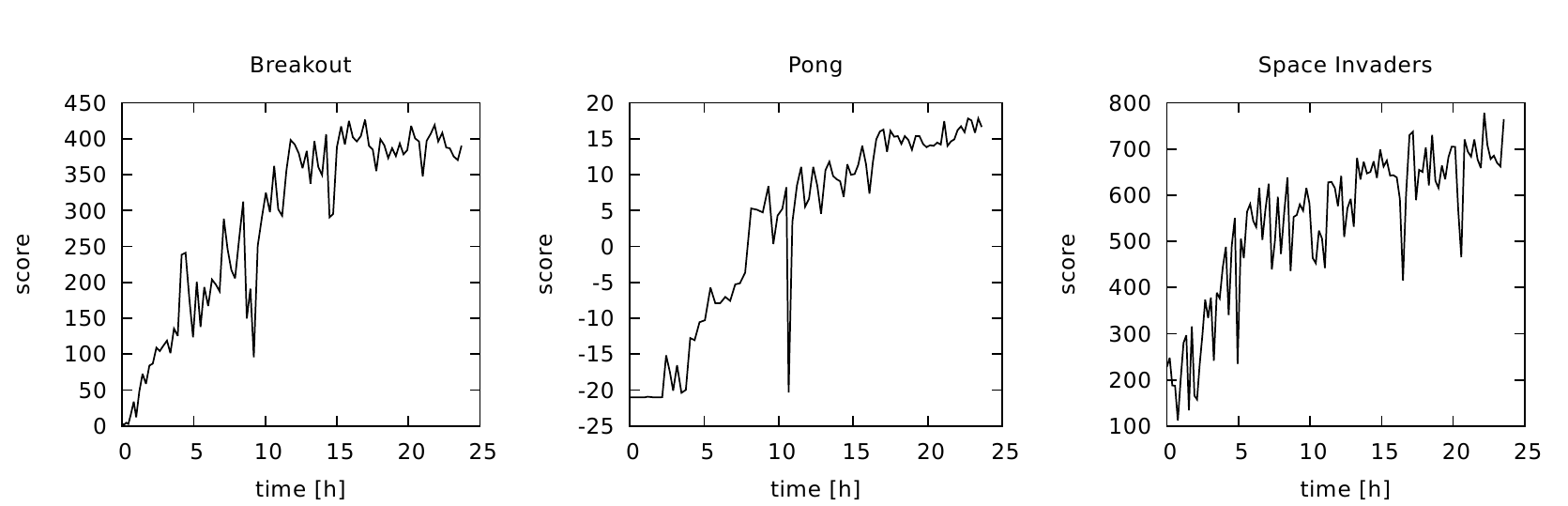}
\caption{Mean score for 50 consecutive games vs training time for the best model obtained for atari Breakout, Pong and Space Invaders.}
\label{plot:score_vs_time}
\end{figure}

By using the custom convolution primitives from the MKL library it was possible to increase the training speed by a factor of 3.8 (from 151.04 examples/s to 517.12 examples/s). This made it possible to train well performing agents in under 24 hours. As a result, novel concepts and improvements to the algorithm can now be tested more quickly, possibly leading to further advances in the field of reinforcement learning. The increase in speed was achieved without hurting the results obtained by the agents trained. Example training curves for 3 different games are presented in the Figure \ref{plot:score_vs_time}.

\subsection{Batch size and learning rate tuning}
\label{subsection:batch}
Using the previously described pipeline optimized for better CPU performance we conducted a series of experiments designed to determine the optimal batch size and learning rate hyperparameters. The experiments were performed using the random search method \cite{bengio_random_search}. For each hyperparameter its value was drawn from a loguniform distribution defined on a range $[10^{-4}, 10^{-2}]$ for learning rate and $[2^1, 2^{10}]$ for batch size. Overall, over 200 experiments were conducted in this manner for 5 different games. The results are presented in the figures \ref{figure:overall},\ref{figure:for_each_game} below. It appears that for the 5 games tested one could choose a combination of learning rate and batch size that would work reasonably well for all of them. However, the optimal settings for specific games seem to diverge.

As one could expect when using large batch sizes, better results were obtained with greater learning rate's. This is most probably caused by the stabilizing effects of bigger batch sizes on the mean gradient vector used for training. For smaller batch sizes using the same learning rate would cause instabilities, impeding the training process.

Overall, batch size of around 32 a learning rate of the order of $10^{-4}$ seems to have been a general good choice for the games tested. The detailed listing of the best results obtained for each game is presented in the Table \ref{table:optimization_results}.
\begin{table}[H]
\centering
\caption{Mean scores and hyperparameters obtained for the best models for each game}
\label{my-label}
\begin{tabular}{|c|c|c|c|c|}
\hline
game              & learning rate      & batch size & score mean & score max \\ \hline
Breakout-v0      & 0.00087 & 22    & 390.28      & 654        \\ \hline
Pong-v0          & 0.00017 & 19    & 16.64       & 21         \\ \hline
Riverraid-v0     & 0.00024 & 87    & 10,018.40   & 11570      \\ \hline
Seaquest-v0      & 0.00160 & 162   & 1,823.41    & 1840       \\ \hline
SpaceInvaders-v0 & 0.00032 & 14    & 764.70      & 2000       \\ \hline
\end{tabular}
\label{table:optimization_results}
\end{table}

\begin{figure}[H]
\centering
\includegraphics[width=1.2\textwidth]{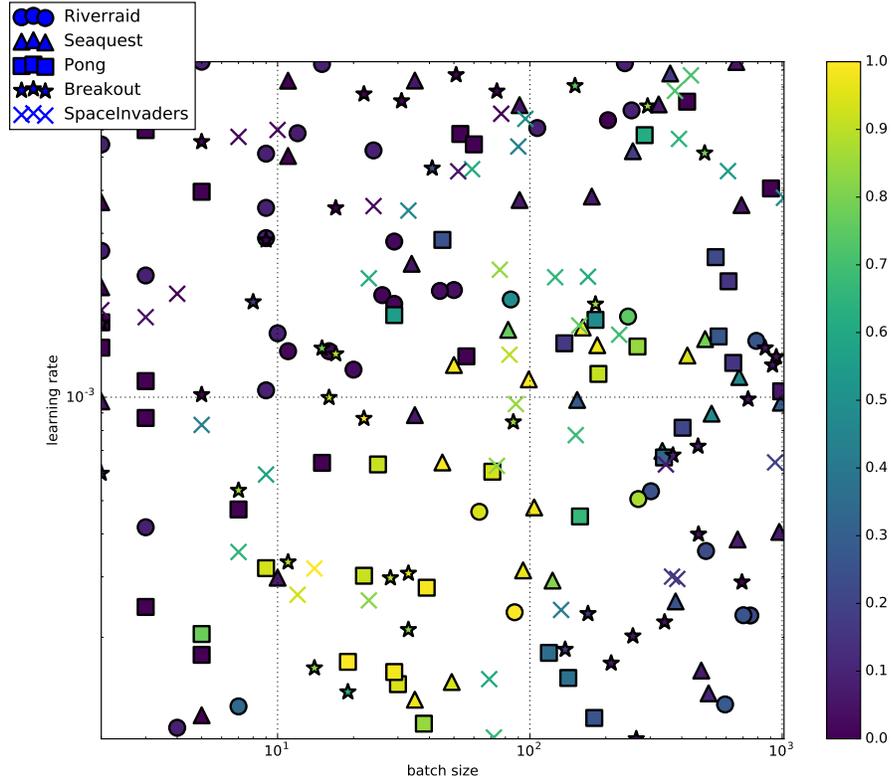}
\caption{Overall results of the random search for all the games tested. The brighter the color the better the result for a given game. Color value 1 means the best score for the game, color value 0 means the worst result for the given game.}
\label{figure:overall}
\end{figure}

\begin{figure}[H]
\hspace*{-1cm}
\includegraphics[width=1.1\textwidth]{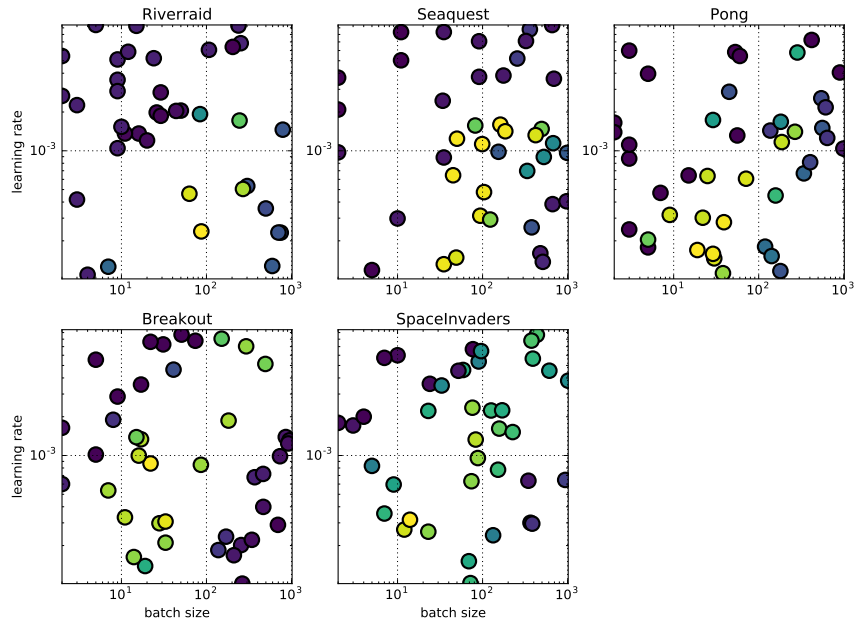}
\caption{Results of random search for each game separately. Brighter colors mean better results.}
\label{figure:for_each_game}
\end{figure}

\section{Conclusions and further work}

Preliminary results contained in this work can be considered as a next step in reducing the gap between CPU and GPU performance in deep learning applications. As shown in this paper, in the area of reinforcement learning and in the context of asynchronous algorithms, CPU-only algorithms already achieve a very competitive performance. 

As the most interesting future research direction we perceive extending results of \cite{allinea} and tuning of performance of asynchronous reinforcement learning algorithms on large computer clusters with the idea of bringing the training time down from hours to minutes. 

Constructing a compelling experiment for the Xeon Phi\textsuperscript{\textregistered}\ platform also seems to be an interesting challenge. Our current approach would require a significant modification because of much slower single core performance of Xeon Phi\textsuperscript{\textregistered}. However, preliminary results on the Pong game are quite promising with a state-of-the-art results obtained in 12 hours on a single Xeon Phi\textsuperscript{\textregistered}\ server.

%ne can expect a better perfoemance 
%We believe that even after the advancements presented above training time can still be decreased substantially. The first approach would be to perform similar tests on stronger CPUs such as the novel Intel Xeon Phi architecture as well as the Skylake processor family. Both of them can execute AVX-512 instructions, the use of which can substantially improve processing speed in compute bound applications such as machine learning.
%An alternative approach would be to utilize the distributed processing paradigm by deploying the algorithm on a large cluster. This can prove a rather challenging task because of the asynchronous nature of the BA3C algorithm. However, if successful, it could enable training of state-of-the art agents for playing Atari games in a matter of minutes, not hours.

\bibliographystyle{splncs03}
\bibliography{bibliography.bib}
\end{document}